\documentclass[%
 aip,
 amsmath,amssymb,
reprint,%
twocolumn,%
]{revtex4-1}

\usepackage{graphicx}% Include figure files
\usepackage{dcolumn}% Align table columns on decimal point
\usepackage{bm}% bold math
\usepackage[utf8]{inputenc}
\usepackage[T1]{fontenc}
\usepackage{mathptmx}

\begin{document}

%\preprint{AIP/123-QED}
\ifpdf
\DeclareGraphicsExtensions{.pdf, .jpg}
\else
\DeclareGraphicsExtensions{.eps, .jpg}
\fi
\def\mymathhyphen{{\hbox{-}}}
\def\hslash{\hbar}
\def\imag{i}
\def\grad{\vec{\nabla}}
\def\div{\vec{\nabla}\cdot}
\def\curl{\vec{\nabla}\times}
\def\DDt{\frac{d}{dt}}
\def\ddt{\frac{\partial}{\partial t}}
\def\ddx{\frac{\partial}{\partial x}}
\def\ddy{\frac{\partial}{\partial y}}
\def\lap{\nabla^{2}}
\def\divv{\vec{\nabla}\cdot\vec{v}}
\def\gradS{\vec{\nabla}S}
\def\vvec{\vec{v}}
\def\wc{\omega_{c}}
\def\<{\langle}
\def\>{\rangle}
\def\Tr{{\rm Tr}}
\def\Csch{{\rm csch}}
\def\Coth{{\rm coth}}
\def\Tanh{{\rm tanh}}
\def\g2{g^{(2)}}
\newcommand{\al}{\alpha}

\newcommand{\la}{\lambda}
\newcommand{\del}{\delta}
\newcommand{\om}{\omega}
\newcommand{\ep}{\epsilon}
\newcommand{\pd}{\partial}
\newcommand{\bra}{\langle}
\newcommand{\ket}{\rangle}
\newcommand{\bbra}{\langle \langle}
\newcommand{\kket}{\rangle \rangle}
\newcommand{\non}{\nonumber}
\newcommand{\be}{\begin{equation}}
\newcommand{\ee}{\end{equation}}
\newcommand{\bea}{\begin{eqnarray}}
\newcommand{\eea}{\end{eqnarray}}

\title{Heat-current filtering for Green-Kubo and Helfand-moment molecular dynamics predictions of thermal conductivity: Application to the organic crystal $\beta$-HMX}
% Force line breaks with \\

\author{Andrey Pereverzev}
\author{Tommy Sewell}
 \email{pereverzeva@missouri.edu.}
  \email{sewellt@missouri.edu.}
\affiliation{Department of Chemistry, University of Missouri-Columbia, Columbia, Missouri 65211-7600, USA}

\date{\today}% It is always \today, today,
             %  but any date may be explicitly specified

\begin{abstract}
%% Text of abstract
The closely related  Green-Kubo and Helfand moment approaches are applied to obtain the thermal conductivity tensor of $\beta$-1,3,5,7-tetranitro-1,3,5,7-tetrazoctane ($\beta$-HMX) 
at $T= 300$ K and $P=1$ atm from equilibrium molecular dynamics (MD) simulations. Direct application of the Green-Kubo formula exhibits slow convergence 
of the integrated thermal conductivity values even for long (120 ns) simulation times. To partially mitigate this slow convergence 
we developed a numerical procedure that involves filtering of the MD-calculated heat current. The filtering is accomplished by
 physically justified removal of  the heat-current component which is given by a linear function of atomic velocities. A double-exponential function is fitted to the 
integrated time-dependent thermal conductivity, calculated using the filtered current, to obtain the asymptotic values for the thermal conductivity. 
In the Helfand moment approach the thermal conductivity is obtained from the rates of change  of the averaged squared Helfand moments.
Both methods are applied to periodic $\beta$-HMX supercells of four different sizes and estimates for the thermal conductivity 
of the infinitely large crystal are obtained using Matthiessen’s rule.
Both approaches yield similar although not identical thermal conductivity values. These predictions are compared to  experimental and other theoretically determined values for the thermal conductivity 
 of $\beta$-HMX.

\end{abstract}
\maketitle
%%Graphical abstract
%\begin{graphicalabstract}
%\includegraphics{grabs}
%\end{graphicalabstract}

%%Research highlights
%\begin{highlights}
%\item Research highlight 1
%\item Research highlight 2
%\end{highlights}

%\begin{keyword}
%% keywords here, in the form: keyword \sep keyword
%thermal conductivity \sep molecular dynamics
%% PACS codes here, in the form: \PACS code \sep code
%\PACS 0000 \sep 1111
%% MSC codes here, in the form: \MSC code \sep code
%% or \MSC[2008] code \sep code (2000 is the default)
%\MSC 0000 \sep 1111
%\end{keyword}

%% \linenumbers

%% main text
\section{Introduction}
Understanding the response of an explosive material subjected to a thermo-mechanical insult is a longstanding challenge to the energetic materials community \cite{Handley,Fried}.  Substantial effort has been devoted to predicting the response of high explosives to shock stimuli, with increased attention in recent years on the development of accurate grain-resolved continuum models and simulations that account explicitly for mesoscale physics and material microstructure as discussed, for example,  in  Ref. \cite{Das}.
%Plastic-bonded explosives (PBXs) are highly heterogeneous materials for which internal interfaces play critical roles in ignition and detonation-initiation processes. Typical grain sizes are on the order of 10-100 micrometers. Therefore, grain-resolved mesoscale descriptions at the representative volume element (RVE) scale and below are a critical step on the development path for predictive, homogenized but physics-based and microstructurally aware, macroscale continuum reactive-burn models.\cite{Perry,Udaykumar,Handley,NicholsIII} However, the mesoscale models can be no more accurate than the fundamental information used in their construction, and therefore require insights and quantitative knowledge of the underlying nanoscale properties and processes. Although experimental data for many needed properties are sparse, much of that information can be obtained from atomic-scale simulation methods.
%

One of the key inputs needed by mesoscale models is the thermal conductivity of the constituent materials. Ignition of chemistry behind a shock wave in energetic materials requires spatial localization of energy, colloquially known as hotspots. For a given hotspot, the competition between local chemical energy release and energy dissipation due to thermal conduction will largely determine whether the hotspot will quench or grow and eventually interact with others, possibly culminating in violent explosion or detonation. This is particularly important in scenarios involving impacts or weak shocks, for which an undesired deflagration-to-detonation transition can lead to disastrous outcomes. Experimentally determined thermal conductivities for explosives are subject to large uncertainties associated with measurement techniques and sample preparation (see, for example, Table I of Ref. \cite{Chitsazi}), and data for elevated pressures and temperatures are particularly rare \cite{Parr,Perriot}. Also, there are precious few experimental determinations of the thermal conductivity tensor for single oriented high-explosive crystals under any conditions \cite{Perriot}. Fortunately, molecular dynamics (MD) simulations are well suited for providing thermal conductivity values and other fundamental information needed by the mesoscale models \cite{Das}.

Calculation of the thermal conductivity tensor for solids and fluids using atomic-scale simulation methods is most easily accomplished using the Green-Kubo (GK) formalism applied in the framework of classical MD. The thermal conductivity tensor $\kappa^{\alpha\beta}$ in the GK approach is given by \cite{Kubo}
\begin{equation}
\kappa^{\alpha\beta}=\frac{1}{k_BT^2V}\int_0^\infty dt\langle J^{\alpha}(0) J^{\beta}(t)\rangle, \label{qcurr}
\end{equation}
where $J^{\alpha}$ is the $\alpha$th Cartesian component of the heat current, $T$ is temperature, $V$ is the system volume, and
the brackets indicate averaging over an equilibrium ensemble.
The GK approach has become a powerful tool for obtaining thermal conductivities of various materials due, in part, to the simplicity of its implementation, which only involves equilibrium MD simulations \cite{Ladd,Che,McGaughey1,McGaughey2,Izvekov,Landry,McGaughey3,Henry,Yip,Fan}; as opposed to ad hoc non-equilibrium methods \cite{MacDonald,Muller,Long,Algaer,KroonblawdSewell,Chitsazi,Perriot,Perriot2}, for which several empirical choices in simulation protocol must be determined or guessed. 
The main disadvantage of the GK approach is a slow convergence of the integral in Eq. (\ref{qcurr}) to its true value. 
This slow convergence may require very long simulation times and careful analysis of the correlation function.\cite{Yip,Chen,McGaughey2}

Another equilibrium MD technique that can be used to obtain thermal conductivity  is the Helfand moment (HM) approach \cite{Helfand,Gaspard}. 
This approach is closely related to the GK method but it has rarely been used for practical calculations of thermal conductivity,
in particular because of the challenge of correctly defining a proper dynamical variable for Helfand moments for periodic systems \cite{Viscardy,Kinaci}.
However, this issue does not arise in the present work because the Helfand moments are expressed as integrals of heat-current components that are properly defined for periodic systems.

In this study, we apply the GK and HM approaches to compute the (300 K, 1 atm) thermal conductivity of  
$\beta$-1,3,5,7-tetranitro-1,3,5,7-tetrazoctane, also known as $\beta$-octahydro-1,3,5,7-tetranitro-1,3,5,7-tetrazocine ($\beta$-HMX), which is an important energetic material that is used in a number of high-performance military 
PBX and propellant formulations \cite{Gibbs}. Several pure polymorphs of HMX are known \cite{CadySmith},  among which $\beta$-HMX is the thermodynamically stable form at standard ambient conditions \cite{Cady}. The thermal properties of $\beta$-HMX are important for understanding processes such as hot spot formation that ultimately lead to ignition and detonation initiation \cite{Das}.  
To alleviate the slow convergence of the thermal conductivity in the GK approach we developed a procedure that involves physically justified filtering of the
heat current prior to calculating the heat-current correlation function, followed by
physically motivated fitting of the time-dependent conductivity to a bi-exponential function.
We study the size dependence of the thermal conductivity tensor using both the GK and HM approaches and estimate its value for infinitely large crystal by extrapolation.
The results are compared to experimental and theoretical values in the literature.

\section{Simulation  Details} \label{sd}
  %\subsection{Force Field and MD Simulation Details} \label{sd}
 All MD simulations were performed using the LAMMPS package \cite{Plimpton}. The nonreactive, fully flexible molecular potential for nitramines proposed by Smith and Bharadwaj \cite{Smith99} and further developed by Bedrov et al. \cite{Bedrov3} and others \cite{Kroonblawd1,Das} was employed. This force field is well-validated and has been used in numerous previous studies of HMX (see, for example, Ref. \cite{Pereverzev} and references therein). In the current study, we used the version described in Ref. \cite{Das}.
% we modified the original nitramine force field used in \cite{Bedrov3} by adjusting the N-O and C-H harmonic bond stretching force constants to better reproduce the corresponding experimental vibrational mode frequencies (see \cite{Kroonblawd1}  for more details). We further modified the original force field by adding a $1/r^{12}$ repulsive-core term to the non-bonded pair interaction potential. The reason for this latter modification is to eliminate the unphysical very-short-range attractive well in the Buckingham pair potential used in the original force field. The parameters of the $1/r^{12}$ repulsive core are chosen in such a way that the system dynamics is practically unaffected under the conditions we study in this paper. More details can be found in the supporting information for the work by Zhao et al.. \cite{Puhan1} A cut-off distance of 11 {\AA} was used for repulsion, dispersion, and short-range Coulomb interactions. Long-range electrostatic interactions were calculated using the PPPM method with the relative error in the forces set to $1\times10^{-6}$.  
 Sample LAMMPS input decks including all force-field parameters, details of how the forces were evaluated, and a crystal supercell description are included in the supplementary material. Simulations were performed for three-dimensionally (3D) periodic supercells of $\beta$-HMX consisting of $4{\bf a}\times4{\bf b}\times4{\bf c}$, $5{\bf a}\times5{\bf b}\times5{\bf c}$, $6{\bf a}\times6{\bf b}\times6{\bf c}$, and $8{\bf a}\times8{\bf b}\times8{\bf c}$ unit cells, where 
 ${\bf a}$, ${\bf b}$, and ${\bf c}$ are the unit-cell lattice vectors in the $P2_1/n$ space group setting. The mapping between the crystal and Cartesian frames is 
 ${\bf{a}}\,\|\,{\hat{\bf{x}}}$, ${\bf{b}}\,\|\,{\hat{\bf{y}}}$, and ${\bf{c}}$ in 
 $+{\hat{\bf{z}}}$ halfspace. Henceforth, we describe system sizes as $m\times m\times m$ for simplicity. The supercell parameters corresponding to $T = 300$ K and $P = 1$ atm were determined as averages over the final 500 ps of 600 ps isobaric-isothermal (NPT) trajectories, sampled every 100 fs, in which all six lattice parameters were allowed to vary independently (LAMMPS {\textit{tri}} keyword). The target stress state was the unit tensor (in atm), the time step was 0.2 fs, and the barostat and thermostat coupling parameters were set to 200 fs and 20 fs, respectively. The resulting lattice parameters are $a = 6.587$ {\AA} , $b = 10.47$ {\AA}, $c = 7.603$ {\AA}, $\beta=98.83^{\circ}$ and $\alpha = \gamma = 90.0^{\circ}$, where the least-significant digit for a given parameter is the last one that was common to all four supercell sizes. The resulting cell parameters were used for 500 ps long 300 K isochoric-isothermal (NVT) trajectories, with velocity re-selections at 100 ps intervals, to prepare the supercells for production trajectories. The NVT time step and thermostat coupling parameter were 0.2 fs and 20 fs, respectively.
 
  The heat current time correlation functions (HCCFs), defined as
  \be
  C^{\alpha\beta}(t)=\langle J^{\alpha}(0) J^{\beta}(t)\rangle,
  \ee
 were obtained from 30 independent isochoric-isoenergetic (NVE) trajectories for each system size. Each trajectory was 4 ns long and the heat-current data were recorded every femtosecond. A time step of 0.1 fs was used.
 Because $\beta$-HMX is a monoclinic crystal, only the following components of the HCCF and the corresponding thermal conductivity tensors are required: $xx$, $yy$, $zz$, $xz$, and $zx$. We show in the supplementary material that the $xy$, $yx$, $yz$,  and $zy$ components of the thermal conductivity tensors are, indeed, numerically zero. For the $xz$ and $zx$ GK HCCF and thermal conductivity tensor components, we report the average of the two components and label it with the superscript $xz$.
 
 The dynamical variable in the GK expression, Eq. (\ref{qcurr}), is the heat current ${\bf J}$. The correct definition of ${\bf J}$ is 
crucial for obtaining accurate values of the thermal conductivity tensor.
 To calculate the heat current, LAMMPS first computes the so-called per-atom stress tensor \cite{Plimpton}.
 The most recent (October 29, 2020) version of LAMMPS at the time the present simulations were performed  can calculate two different types of per-atom stress tensor, specified by 
 using the  {\it centroid/stress/atom} and {\it stress/atom} keywords. The two definitions of the per-atom stress are identical 
 for two-body 
 interaction potentials but differ for potentials involving higher-order interaction terms. Recently, it was shown theoretically and numerically  
 that the per-atom stress obtained using the {\it stress/atom} keyword does not give correct values of the heat current
 for systems with three- and four-body interactions, and that the {\it centroid/stress/atom} keyword should be used instead \cite{Surblys}.  
 However, the  version of LAMMPS used here does not support the {\it centroid/stress/atom} keyword  
 for potentials with long-range Coulombic interactions, which are present in the force field. 
 
 To overcome this limitation we used the 
 following hybrid approach to calculate the heat current: The {\it stress/atom} keyword was used for the part of the current arising from  all two-body interactions 
 (including the Coulombic ones) and the {\it centroid/stress/atom} keyword was used for the part of the current due to three- and four-body interactions, 
 viz. angles, dihedrals, and improper dihedrals. The two parts were then added to obtain the total heat current, which we will refer to as the hybrid heat current. This current was used to 
 obtain 
thermal conductivities reported in Sec. \ref{rad}.  A comparison of these results to ones obtained  using the {\it stress/atom} keyword alone  is also provided.

\section{Results, analysis, and discussion} \label{rad}

\subsection{Heat current filtering}
\begin{figure*}
 \includegraphics[width=1.0\textwidth]{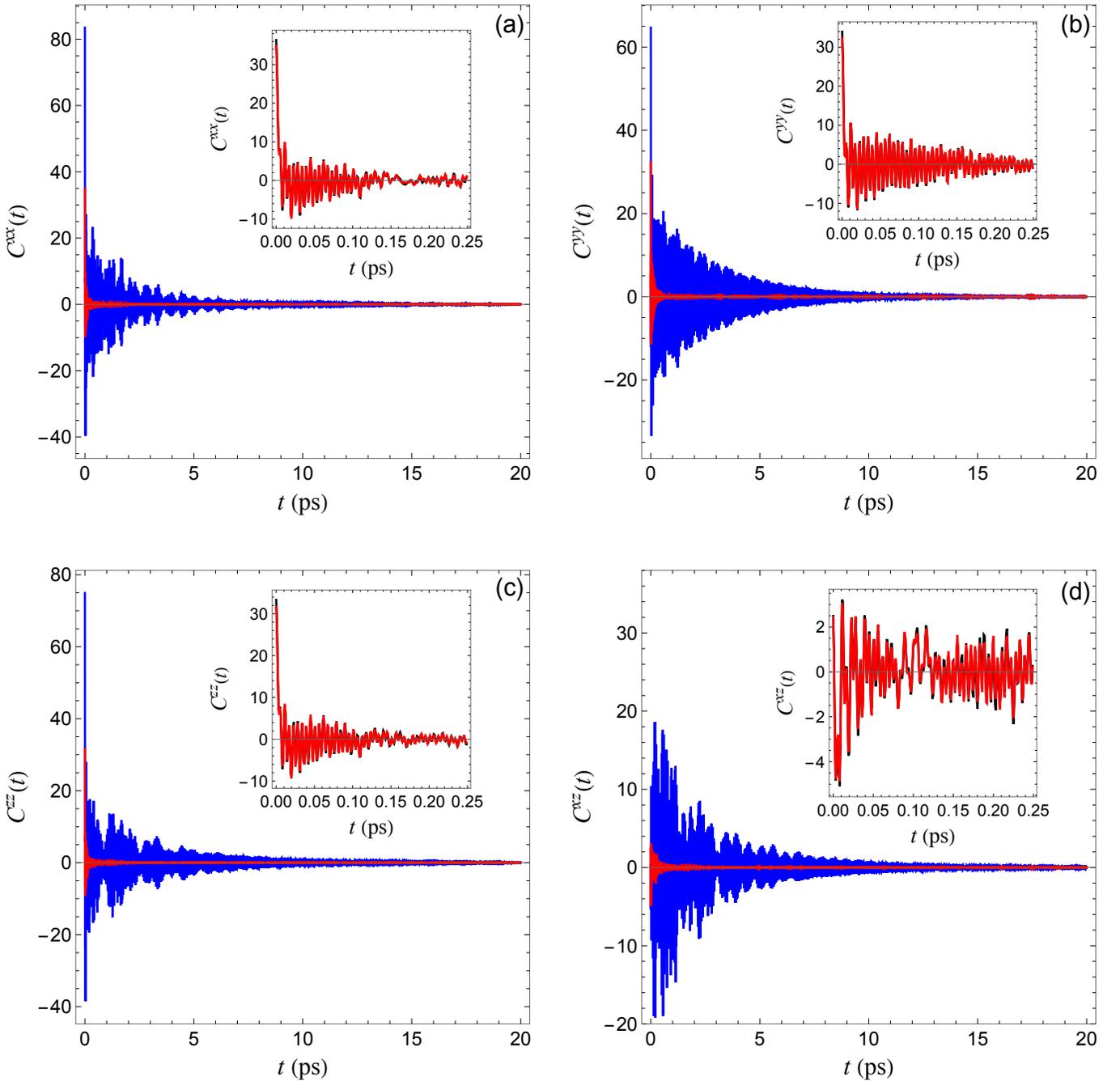}  
 \caption{\label{Figure1}  (a)  The heat current correlation function  $C^{xx}(t)$ for the $4\times 4\times 4$ supercell  in units of 
 ${\textrm{kcal}}^2\,{\textrm{fs}}^2\,{\textrm{mol}}^{-2}\,$\AA$^{-2}$,
 calculated using the unfiltered hybrid heat current computed by LAMMPS (blue) 
 and the hybrid heat
 current filtered using the second scheme (red). The inset compares $C^{xx}(t)$ over the first 0.25 ps obtained using 
  the first (black) and second (red) filtering schemes.
 Panels (b), (c), and (d): Same as (a) but for $C^{yy}(t)$, $C^{zz}(t)$, and $C^{xz}(t)$, respectively.}
 \end{figure*}

\begin{figure*}
 \includegraphics[width=1.0\textwidth]{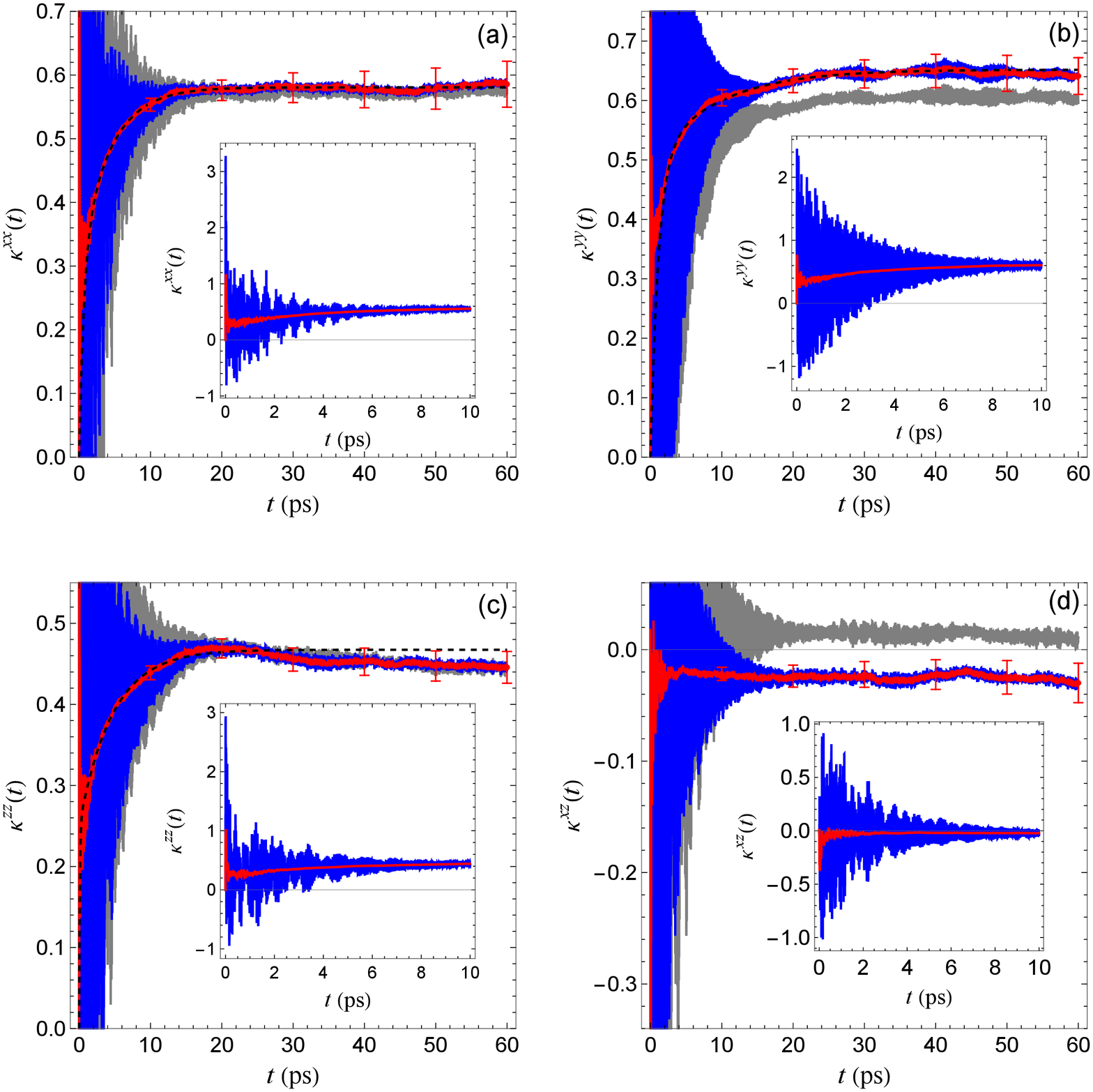}  
 \caption{\label{Figure2}  (a) Time-dependent thermal conductivity tensor component $\kappa^{xx}(t)$  for the $4\times 4\times 4$ supercell  in units of $\textrm {W\,m}^{-1}\,\textrm{K}^{-1}$, obtained using the unfiltered hybrid heat current calculated by LAMMPS (blue) 
 and the filtered hybrid heat 
 current from the second filtering scheme (red). Uncertainties for the red curve reflect one-sigma standard error. The dashed black curve is the double-exponential function fitted to the red curve between 2 ps and 30 ps. The gray curve shows $\kappa^{xx}(t)$ when the LAMMPS {\it stress/atom} keyword alone is used to calculate the heat current with no filtering. 
 The inset shows the full range of $\kappa^{xx}(t)$ over the first 10 ps.
Panels (b), (c), and (d): Same as (a) but for $\kappa^{yy}(t)$, $\kappa^{zz}(t)$, and $\kappa^{xz}(t)$, respectively. No fitting function was used for 
 $\kappa^{xz}(t)$.
 }
 \end{figure*}

Typical HCCFs for $\beta$-HMX at (300 K, 1 atm) obtained using the hybrid heat current calculated by LAMMPS  are shown in blue in Fig. \ref{Figure1}.
One can see that the correlation functions decay in a highly oscillatory manner. Qualitatively similar behavior of the HCCFs was observed for other
polyatomic crystals such as
  $\alpha$-quartz \cite{McGaughey2} and $\alpha$-1,3,5-trinitro-1,3,5-triazinane ($\alpha$-RDX) \cite{Izvekov}. The corresponding time-dependent thermal conductivities, 
 defined as 
 \be
 \kappa^{\alpha\beta}(t)=\frac{1}{k_BT^2V}\int_0^t d\tau\langle J^{\alpha}(0) J^{\beta}(\tau)\rangle, \label{tkappa}
 \ee
 are shown in blue in Fig. \ref{Figure2}. Note that $\kappa^{\alpha\beta}$ in Eq. (\ref{qcurr}) can be viewed as $\lim_{t\to\infty}\kappa^{\alpha\beta}(t)$.
As for the HCCFs in Fig. \ref{Figure1}, the  $\kappa^{\alpha\beta}(t)$ in Fig. \ref{Figure2} are also highly oscillatory over approximately the first 20 ps.   
 It has been understood for some time that at least some of the oscillations in the HCCFs do not contribute to the thermal conductivity because the time integrals over such oscillations vanish \cite{Landry}.
Recently, these observations were given more rigorous theoretical footing. In particular, it was shown \cite{Baroni,Baroni2}
that  heat-current definitions which differ by the time derivative of a 
bounded function of time yield identical thermal conductivity tensors. Expressions for the heat current that differ by such 
time derivatives can be thought of  as different gauges of the heat current \cite{Baroni,Baroni2}. 

Here, we show how some of the oscillations in the HCCFs can be eliminated by physically justified filtering of the heat current. This leads to reduced oscillations in $\kappa^{\alpha\beta}(t)$ as well 
and allows one to apply more rigorous fittings to obtain thermal conductivities.

The heat current for a system consisting of  $N$ atoms as calculated by LAMMPS is defined by the following expression:
\be
{\bf J}=\sum_{i=1}^N\epsilon_i{\bf v}_i-\sum_{i=1}^N{\bf S}_i\cdot{\bf v}_i, \label{curdef}
\ee
where $\epsilon_i$  is the energy of atom $i$ and ${\bf S}_i$ is the per-atom stress tensor of that atom \cite{Plimpton}. 
The tensor ${\bf S}_i$ in Eq. (\ref{curdef}) multiplies
atomic velocity ${\bf v}_i$ as a $3\times3$ matrix multiplies a vector, to yield a vector. The atomic energy $\epsilon_i$ is given by  
\be
\epsilon_i=\frac{m_i |{\bf v}_i |^2}{2}+u_i, \label{locen}
\ee
where the first and second terms on the right-hand side are, respectively, the atomic kinetic and potential energies.
There is a well-known ambiguity in defining $u_i$ for any system with interatomic interactions \cite{Hardy, Allen,Schelling}. 
In this work we use $u_i$ as defined in LAMMPS \cite{Plimpton}.

When analyzing coordinate-dependent dynamical variables in solids it is customary to expand them in 
a Taylor series of displacements of atomic coordinates about their values at the minimum-energy 
configuration and truncate this expansion at the cubic or quartic terms. This procedure is used to convert these dynamical variables to the normal-mode picture. In the case of the heat current, such expansion amounts to expanding $u_i$ and ${\bf S}_i$ 
in Eq. (\ref{curdef}) in a Taylor series of atomic displacements as was done by Hardy \cite{Hardy}.  
With the definition of $u_i$ chosen by Hardy in his theoretical analysis \cite{Hardy}, such expansions for $u_i$ and ${\bf S}_i$ start,
respectively, with 
quadratic and linear terms. 
By contrast, the expansions
for $u_i$ and ${\bf S}_i$ for typical Hamiltonians 
used in MD start with zeroth-order terms,  $u^0_i$ and ${\bf S}^0_i$, which represent, respectively, the atomic potential energy and per-atom stress evaluated 
at the energy-minimized system geometry. Thus, the lowest-order terms in the MD-relevant heat-current expansion are linear in atomic velocities and do not depend on atomic coordinates. These terms 
do not contribute to the thermal conductivity tensor obtained using Eq. (\ref{qcurr}) because they represent the time derivative of a bounded function of time. Indeed,  each Cartesian component of velocity $v^{\alpha}_i$ of atom $i$  is given by
$v^{\alpha}_i=\frac{d r^{\alpha}_i}{d t}$, where $r^{\alpha}_i$ is the $\alpha$th Cartesian component of the position vector of that atom. The atomic coordinates in solids are bounded 
functions of time because atoms are moving in the vicinity of their equilibrium positions. 

When transformed to the normal-mode picture, the linear velocity terms map into terms which are linear in the normal-mode momenta and manifest as decaying oscillations in the HCCFs. It is easy to
 verify that for crystalline solids these linear normal-mode momenta terms involve only optical modes with zero wave vector, and that for crystals with $n$ particles in the unit cell there are at most $3n-3$ such modes.

It is interesting to note that although the linear velocity terms are present in the heat current for generic MD Hamiltonians, they effectively vanish in monatomic crystals with certain high-symmetry lattices; examples include argon \cite{Ladd,McGaughey1}, diamond \cite{Che},  silicon \cite{Henry}, and germanium \cite{Dong2}. This is because for such crystals the $u^0_i$'s are identical for different $i$ values and thus, because these crystals are monatomic,  $\sum_{i}u^0_i{\bf v}_i$ is proportional to the total linear momentum of the system, an integral of motion that is typically set to zero in MD simulations. Similarly, for such high-symmetry monatomic crystals the ${\bf S}^0_i$'s are also identical for different $i$'s and are directly proportional to the $3\times3$ identity matrix, so  $\sum_{i}{\bf S}^0_i\cdot{\bf v}_i$ is also proportional to total linear momentum of the system. These facts explain, at least in part, why the HCCFs of such  systems are much less oscillatory than those of polyatomic crystals \cite{McGaughey2,Izvekov}. (The HCCFs for simple monatomic crystals do, however, exhibit oscillations \cite{Ladd,McGaughey1,Che,Henry,Dong2}. Although oscillations are not predicted for any crystal using the Peierls heat current \cite{Peierls}, we showed recently \cite{PereverzevSewell} that they follow naturally for all physically realistic crystals if the Peierls expression is extended  by taking into account all quadratic terms in the expansion of the Hamiltonian on which it is based rather
 than only the diagonal ones.) 

Based on the preceding discussion, we consider two schemes for eliminating linear velocity terms from the heat current. They are conceptually different but lead to very similar results numerically.
In the first scheme we eliminate the linear terms in the Taylor expansion of the heat current; that is, we subtract $\sum_{i=1}^Nu^0_i{\bf v}_i-\sum_{i=1}^N{\bf S}^0_i\cdot{\bf v}_i$ from the heat current calculated by LAMMPS.

In the second scheme, first proposed in Ref. \cite{Baroni2}, one adds to the heat current a  function of the form  $\sum_{i{\beta}}c_i^{\alpha\beta}v^{\beta}_i$, where the $c_i^{\alpha\beta}$ are adjustable parameters, and  seeks  $c_i^{\alpha\beta}$ which minimize the HCCF
at $t=0$; that is, one finds the adjusted current $\tilde J^{\alpha}= J^{\alpha}+\sum_{i{\beta}}c_i^{\alpha\beta}v^{\beta}_i$ such that $\bra \tilde J^{\alpha}\tilde J^{\alpha}\ket$ is a minimum as a function of the parameters $c_i^{\alpha\beta}$.
Using the definition of the heat current, Eq. (\ref{curdef}), one obtains after some calculations, 
\be
c_i^{\alpha\beta}=-\frac{5kT}{2}-\bra u_i \ket +\bra S_i^{\alpha\beta} \ket. \label{css}
\ee
Thus, in the second scheme  we subtract $\sum_i^N(5kT/2+\bra u_i \ket){\bf v}_i-\sum_i^N\bra{\bf S}_i\ket\cdot{\bf v}_i$ from the heat current calculated by LAMMPS.
The last expression differs from the one used in the first scheme by the presence of the $5kT/2$ term and the use of 
ensemble-averaged $u_i$ and ${\bf S}_i$
rather than $u^0_i$ and ${\bf S}^0_i$.

The main panels of Fig. \ref{Figure1} show HCCF components calculated using the hybrid heat current given by LAMMPS
(blue, see Sec. \ref{sd}) in comparison to the same components calculated using the filtered hybrid current from the second scheme (red). The insets in Fig. \ref{Figure1}  compare results for the two current-filtering schemes; black and red for the first and second schemes, respectively. 
It is obvious from the insets that the HCCF components for
the filtered currents are very similar and, from the main panels, that they exhibit considerably lower amplitude oscillations compared to the ones obtained from the  LAMMPS current. Mathematically, the similarity of the two filtering schemes stems from the facts that $u_i^0\approx \bra u_i \ket$, ${\bf S}_i^0\approx \bra {\bf S}_i \ket$, and
the $5kT/2$ term in Eq. (\ref{css}) is much smaller than the maximum values of the other two terms.  Although not obvious from the insets, the second current-filtering scheme gives slightly 
smaller oscillations compared to the first scheme. Because of this we use the second scheme in the subsequent analysis. 

The corresponding time-dependent thermal conductivities are shown in Fig. \ref{Figure2}.
One can see that using the filtered current (red) drastically reduces the amplitude of oscillations in the integrals compared to the case of 
no filtering (blue). The black dashed curves are discussed in the next subsection and the gray curves in Sec. \ref{perdef}. 
\subsection{Time-dependent heat conductivity fitting}
A serious challenge when applying the GK approach is the slow convergence of the time-dependent thermal conductivity to a constant value.
Even using a total simulation time of 120 ns to obtain the HCCF, the time-dependent thermal conductivity is not fully converged to a constant value, as can be seen in Fig. \ref{Figure2} for both the unfiltered and filtered versions of the conductivity. A similar slow convergence of the thermal conductivities has been observed 
in GK MD simulations for other solids as well \cite{Yip,McGaughey1,McGaughey2,Landry,Chen,Fan}.

Several approaches that deal with the slow thermal conductivity convergence have been proposed, all of which are based on
analysis of the HCCF. The ``first-dip method'' proposed by Li et al. \cite{Yip}  uses the time 
for which the HCCF first reaches zero as the time at which the time-dependent thermal conductivity should be evaluated;
that is, as the upper limit to the integral in Eq. (\ref{tkappa}).
However, this method cannot be applied for polyatomic solids because of the oscillatory behavior of the correlation functions even for short times
when the function envelope  is still very far from being small.
A more rigorous approach was proposed by Chen et al. \cite{Chen}, who argued that the slow convergence of the time-dependent thermal conductivity 
is due to the HCCF being not reliable for long times because it is calculated from a finite-length simulation trajectory. They proposed to truncate the HCCF 
after the  time when the ratio of standard deviation for the correlation function to the correlation function itself first exceeds 
unity. Unfortunately, this approach cannot be applied in our case, again because of the oscillatory nature of the HCCF.
Finally, another approach, often used in the literature \cite{Che,McGaughey1,McGaughey2,Izvekov}, is to assume 
a certain functional form for the correlation function and then fit that 
function to the HCCF data. Then, the thermal conductivity is obtained by integrating the fitted function from zero to infinity.
The function that is most commonly used for fitting of HCCFs for  monatomic crystals is the double exponential 
of the form \cite{Che,McGaughey1}
\be
f(t)=A_{1}\exp(-\gamma_{1} t)+A_{2}\exp(-\gamma_{2} t). \label{fitting}
\ee
Physically, the first and second terms in Eq. (\ref{fitting}) account approximately for the correlation function decay due to 
the relaxation of the acoustic phonons with long and short wavelengths, respectively \cite{McGaughey1}.
For polyatomic crystals, the correlation functions are usually highly oscillatory (as in the present study) and the suggested fitting functions
are given by sums of exponentials and cosines modulated by exponentials \cite{McGaughey2,Izvekov}.
However, as discussed above, a significant part  of the oscillations in the correlation functions does not contribute to the
thermal conductivity because the time integral over them vanishes. Therefore,  fitting oscillatory functions to the HCCF data may lead to incorrect estimates of the thermal conductivity
because doing so involves fitting functions with finite time integrals  to the data, some parts of which have vanishing time integrals.

As can be seen from Fig. \ref{Figure2}, filtering of the heat current does not improve 
the asymptotic convergence of the time-dependent thermal conductivities. However, by eliminating a considerable part of oscillations, the filtering exposes a much smoother behavior for the diagonal components
of the time-dependent thermal conductivities  during the time interval 
when the conductivities undergo substantial growth ($t\lesssim 20$ ps), and this motivates the following alternative fitting procedure to obtain the conductivity.

In view of the discussion in the preceding paragraphs and the general form of the time-dependent thermal conductivity obtained from the filtered current (see Fig. \ref{Figure2}),  rather than using fittings to the correlation functions we propose to use fitting to the time-dependent
thermal conductivities. We use the fitting function  
\be
F(t)=B_1+B_2-B_1\exp(-\gamma_1 t)-B_2\exp(-\gamma_2 t),
\ee
which is obtained by integrating the function $f(t)$ in Eq. (\ref{fitting}) over time. The asymptotic value of the thermal conductivity
is then given by $B_1+B_2$. We chose to fit the time-dependent thermal conductivity data in the interval between $t=2$ ps and 30 ps 
because in this region most data for the time-dependent thermal conductivity exhibit monotonic increases. Fitting to the double-exponential approach is 
applicable only 
for the diagonal components of the thermal conductivity tensor. The off-diagonal component $\kappa^{xz}(t)$ is much smaller in absolute value, much noisier, and
does not exhibit a clear exponential-type behavior, as can be seen in Fig. \ref{Figure2}d. Therefore, for the case of $\kappa^{xz}(t)$ 
we will report the values of the integrals taken at 10 ps, the time for which $\kappa^{xz}(t)$ becomes approximately constant while its
error bars are still much smaller than the absolute value of $\kappa^{xz}(t)$.
 
\subsection{Helfand moment approach}
\begin{figure}
 \includegraphics[width=\columnwidth]{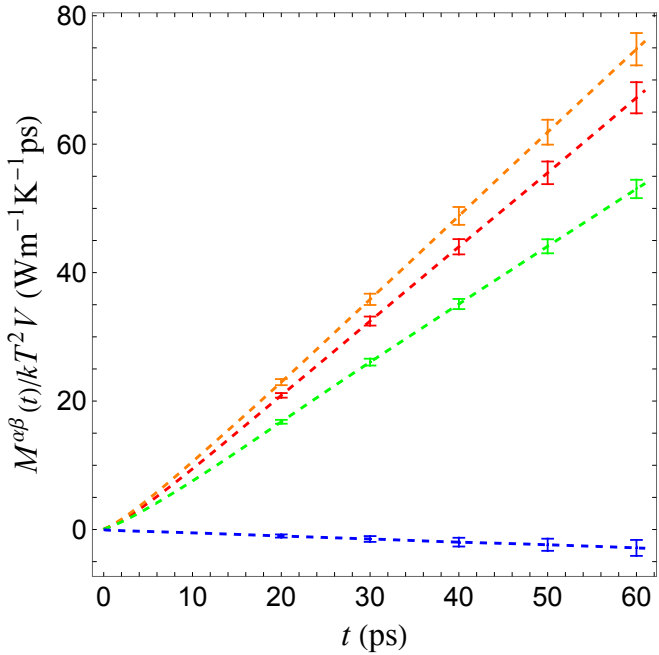}  
 \caption{\label{Figure3}  Functions $M^{xx}(t)$ (red), $M^{yy}(t)$ (orange), $M^{zz}(t)$ (green), and $M^{xz}(t)$ (blue) for the  
 $4\times 4\times 4$ supercell, calculated using the hybrid heat current and second filtering scheme. Uncertainties corresponding to one-sigma standard error  
 are shown for selected times.}
 \end{figure}

\begin{figure*}
 \includegraphics[width=1.0\textwidth]{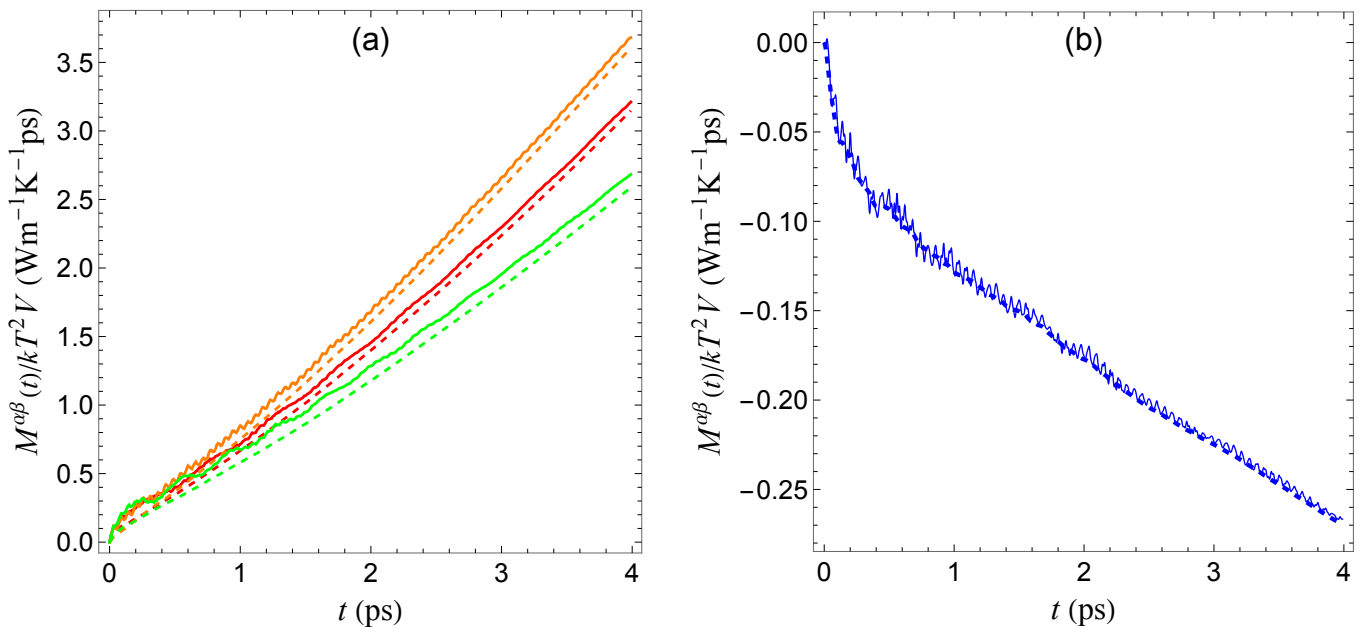}  
 \caption{\label{Figure3prime}  (a) Functions $M^{xx}(t)$ (red), $M^{yy}(t)$ (orange), and $M^{zz}(t)$ (green) for the first 4 ps for the
 $4\times 4\times 4$ supercell calculated using filtered (dashed curves) and unfiltered (soild curves) currents. Panel (b): Same as (a) but for function $M^{xz}(t)$.}
 \end{figure*}

Because of the challenges encountered when using the brute-force GK method, we also considered an alternative analysis  based on the closely related Helfand moment approach \cite{Helfand,Gaspard}, which follows directly from the GK expression.
In this approach, the Helfand moment $G^{\alpha}(t)$ is defined in terms of the time integral of the heat-current component $J^{\alpha}$ as 
\be
G^{\alpha}(t)=G^{\alpha}(0)+\int_0^tJ^{\alpha}(\tau)d\tau. \label{hemo}
\ee
One then evaluates the ensemble-averaged quantity
\be
M^{\alpha\beta}(t)=\left\bra \big( G^{\alpha}(t)-G^{\alpha}(0)\big)\big(G^{\beta}(t)-G^{\beta}(0)\big) \right\ket.
\ee
It can be shown \cite{Gaspard} that 
\be
\lim_{t\to\infty}M^{\alpha\beta}(t)=2tk_BT^2V\kappa^{\alpha\beta},
\ee
that is, for long times $M^{\alpha\beta}(t)$ grows linearly as a function of time
with a slope that is directly proportional to the corresponding thermal conductivity tensor component. Since the HM approach is based on the GK formula it may 
exhibit the same slow convergence issues as the GK approach, with long simulation
times required to achieve a constant slope for functions $M^{\alpha\beta}(t)$.
To calculate the Helfand moments (Eq. \ref{hemo}) we used both the unfiltered hybrid heat  current and the filtered hybrid heat  current,  obtained using the second filtering scheme, and integrated these currents numerically using the trapezoidal rule.
Examples of functions $M^{\alpha\beta}(t)$ for the filtered current are shown in Fig. \ref{Figure3}.
(Although not shown, the functions $M^{\alpha\beta}(t)$ obtained using the unfiltered current exhibit slopes that are essentially identical to those shown for the filtered current in Fig. 3, for $t\gtrsim 10$ ps.) Figure   \ref{Figure3prime}
shows functions $M^{\alpha\beta}(t)$ for the first 4 ps, calculated using the filtered and unfiltered hybrid currents. The color scheme is the same as in Fig. 
\ref{Figure3}; results for the unfiltered case are shown as solid curves. It can be seen that functions $M^{\alpha\beta}(t)$ calculated using the unfiltered current exhibit some oscillations, but the amplitude of these oscillations decreases with time and the slopes of $M^{\alpha\beta}(t)$ for the filtered and unfiltered currents become virtually identical. We used functions $M^{\alpha\beta}(t)$ calculated from the filtered current in subsequent analysis of thermal conductivity.
% \begin{figure}
% \includegraphics[width=\columnwidth]{h1.eps}  
% \caption{\label{Figure3}  (a) Function $M^{xx}(t)$ for crystal cells consisting
% of $4\times 4\times 4$ (red),  $5\times 5\times 5$ (orange), $6\times 6\times 6$ (green), and $8\times 8\times 8$ (blue). Error bars for selected times 
% are also shown.
% (b) Same as (a) but for $M^{yy}(t)$. (c) Same as (a) but for $M^{zz}(t)$. (d) Same as (a) but for $M^{xz}(t)$.}
% \end{figure}
The diagonal HM conductivity tensor components $\kappa^{\alpha\alpha}(t)$ were obtained from the slope of a straight line fitted to 
the  $M^{\alpha\alpha}(t)$ data in the interval from $t=30$ ps to 60 ps. In this interval  the $M^{\alpha\alpha}(t)$ exhibit 
essentially linear behavior while still having small relative standard errors. For $M^{xz}(t)$ we used the interval from 10 ps to 20 ps because 
for this component the standard error grows quickly with time and becomes similar in magnitude to  $M^{xz}(t)$ itself after about 30 ps.
Note that both of the fitting intervals used for the  HM approach are outside  the interval used for the analysis of the diagonal components in the GK approach.
\subsection{Size dependence of thermal conductivity}
The GK and HM  approaches were used to  calculate
the thermal conductivity tensor for the $4\times4\times4$, $5\times5\times5$, $6\times6\times6$, and $8\times8\times8$ supercells, to
study the size dependence.
%The time-dependent thermal conductivities obtained from the adjusted heat current and the corresponding fitting curves for GK approach are shown in Figure 
%\ref{Figure3}.
% \begin{figure}
% \includegraphics[width=\columnwidth]{SIZES.eps}  
% \caption{\label{Figure3}  (a) $\kappa^{xx}(t)$ for crystal cells consisting
% of $4\times 4\times 4$ (red),  $5\times 5\times 5$ (orange), $6\times 6\times 6$ (green), and $8\times 8\times 8$ (blue)  
% calculated using the adjusted heat current. The fitting curves are shown, respectively, in black dashes, black, brown dashes, and brown.
% (b) Same as (a) but for $\kappa^{yy}(t)$. (c) Same as (a) but for $\kappa^{zz}(t)$. (d) Same as (a) but for $\kappa^{xz}(t)$. No fitting curves
% are shown.}
% \end{figure}
The conductivities obtained from both approaches are reported in
Table \ref{t1}. 

\begin{figure*}
 \includegraphics[width=1.0\textwidth]{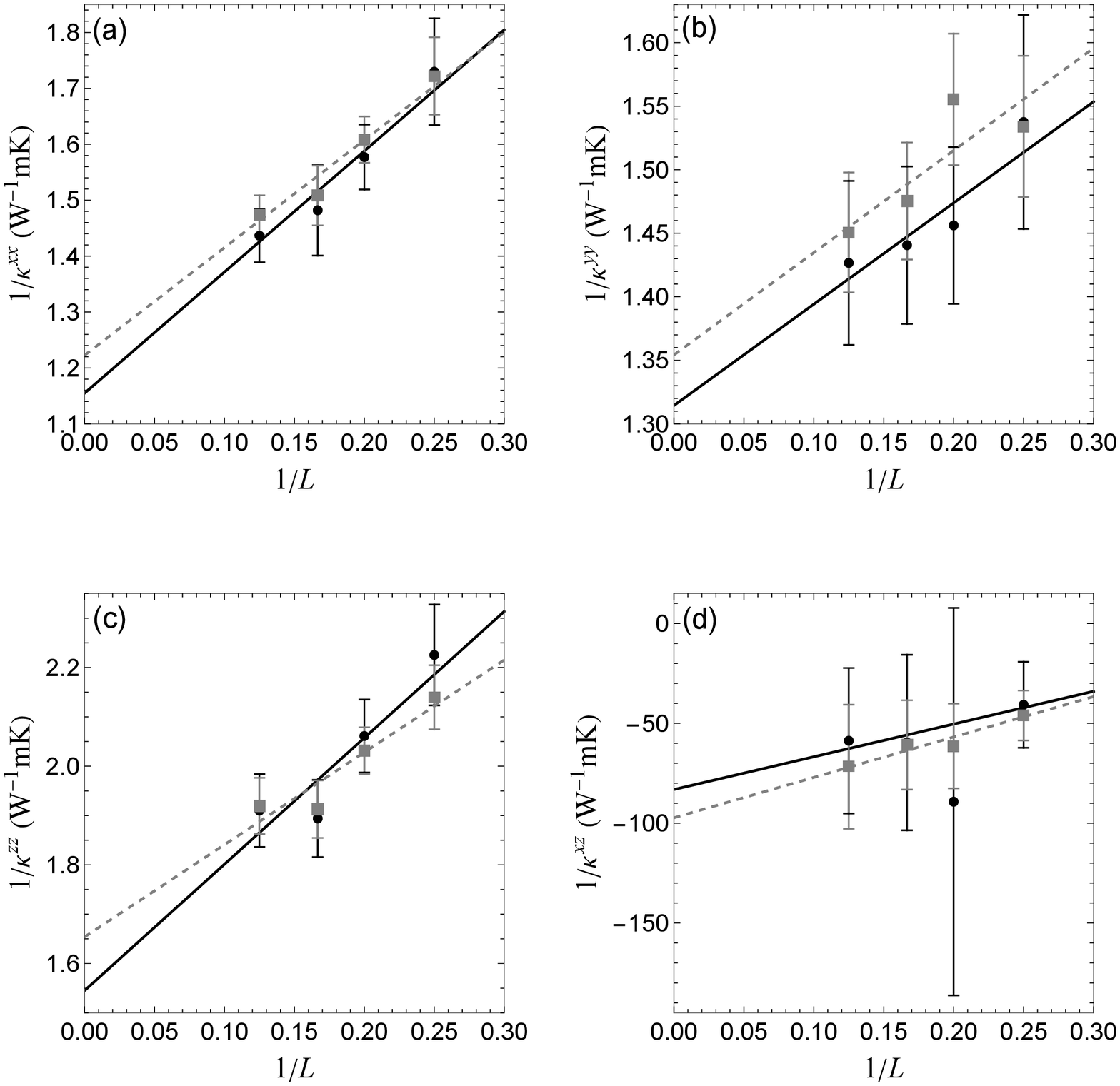}  
 \caption{\label{Figure4}  (a) $1/\kappa^{xx}$  as a function of the inverse crystal supercell length $1/L$ (in arbitrary units) obtained for four different supercell sizes using the  GK
 (gray squares) and HM (black circles) approaches, calculated using the hybrid heat current and second filtering scheme. Uncertainties correspond to one-sigma standard error.
 The dashed gray (GK) and black (HM) lines are
 fits to the data obtained using Eq. (\ref{Matth}).
 Panels (b), (c), and (d): Same as (a) but for $1/\kappa^{yy}$, $1/\kappa^{zz}$, and $1/\kappa^{xz}$, respectively.}
 \end{figure*}

The thermal conductivity tensor of bulk $\beta$-HMX is obtained using Matthiessen’s rule   \cite{Berman,Chern}, 
\be
\frac{1}{\kappa^{\alpha\beta}(L)}=\frac{1}{\kappa^{\alpha\beta}(\infty)}+\frac{A}{L}, \label{Matth}
\ee
where $L$ is the linear size of the crystal and $A$ is a constant.  Matthiessen’s rule is an empirical relation 
between the thermal conductivity and the crystal size. The rule appears to adequately describe the size dependence of thermal conductivity 
of 3D-periodic molecular crystals calculated using  direct methods \cite{Kroonblawd3,Chitsazi,Perriot2},  in which a thermal gradient is imposed and the resulting heat current calculated (or vice versa), with the conductivity evaluated using Fourier's law. Thus, we assume it is also applicable when the thermal conductivity of 3D-periodic crystals is calculated using the GK and HM methods.
Lines  fitted to the size-dependent GK and HM thermal conductivities are shown in Fig. \ref{Figure4}. Although there is some scatter, uncertainties for the GK and HM predictions overlap in all cases. The infinite-size limits of the thermal conductivity tensor components $\kappa^{\alpha\beta}(\infty)$ are reported
 at the bottom of Table \ref{t1}.
 
\begin{table*}
\caption{Thermal conductivity tensor components of  $\beta$-HMX for different supercell sizes obtained from GK and HM analyses, based on the hybrid heat current and second filtering scheme. Units are  $\textrm {W\,m}^{-1}\,\textrm{K}^{-1}$. One-sigma uncertainties are reported based on the approach discussed in the supplementry material.}
\centering
%% \tablesize{} %% You can specify the fontsize here, e.g., \tablesize{\footnotesize}. If commented out \small will be used.
\begin{tabular}{ccccccccccc}
%\toprule
\textrm{Cell size}	&\textrm{Method}&$\kappa^{xx}$	&&$\kappa^{yy} $&&$\kappa^{zz}$&&$\kappa^{xz}$& & $\bar\kappa$ \\
\hline
$4\times4\times4$		&GK& $0.58\pm0.02$ &&$ 0.65\pm0.02 $ && $0.47\pm0.01$ && $-0.022\pm0.006$ &&$0.57\pm0.01$\\
		&HM& $0.58\pm0.03$ &&$ 0.65\pm0.04 $ && $0.45\pm0.02$  && $-0.02\pm0.01$&&$0.56\pm0.02$\\
\hline
$5\times5\times5$		&GK&$0.62\pm0.02$  &  &  $0.64\pm0.02$  &&$0.49\pm0.01$ && $-0.016\pm0.006$ && $0.58\pm0.01$  \\
&HM		&$0.63\pm0.02$   &&$0.69\pm0.03$ &&$0.49\pm0.02$&&$-0.01\pm0.01$&& $0.60\pm0.01$  \\
\hline
$6\times6\times6$		&GK&$0.66\pm0.02$ & &$0.68\pm0.02$ &&$0.52\pm0.02$& &$-0.016\pm0.006$&& $0.62\pm0.01$ \\
&HM &		$0.67\pm0.04$ & &$0.69\pm0.03$ &&$0.53\pm0.02$&& $-0.02\pm0.01$&& $0.63\pm0.02$ \\
\hline
$8\times8\times8$	&GK&$0.68\pm0.02$ &&$0.69\pm0.02$&& $0.52\pm0.02$&& $-0.014\pm0.006$&& $0.63\pm0.01$\\
&HM &	$0.70\pm0.02$ &&$0.70\pm0.03$&& $0.52\pm0.02$&& $-0.02\pm0.01$&&$0.64\pm0.01$\\
\hline
$\infty$	&GK&$0.82\pm0.06$  &&$0.74\pm0.06$&& $0.60\pm0.05$ && $-0.010\pm0.005$&&$0.72\pm0.03$ \\
&HM	&$0.87\pm0.09$ &&$0.76\pm0.08$&&$0.65\pm0.07$ && $-0.01\pm0.01$&& $0.76\pm0.05$ \\
\hline
\end{tabular}
\label{t1} 
\end{table*}
The average value of the thermal conductivity, formally applicable to a polycrystalline sample if interfacial effects are neglected, was obtained as one third of the trace of the thermal conductivity tensor and is 
included in the right-most column of Table \ref{t1}. Note that although small, the nonzero (negative) values of $\kappa_{xz}$ imply that two principal axes of the thermal conductivity tensor are not aligned with the $x$ and $z$ axes in the $xz$ plane. This can be expected for a monoclinic crystal.
\subsection{Comparison of thermal conductivities for different definitions of per-atom stress} \label{perdef}
As discussed in Sec. \ref{sd}, the heat current calculated using the {\it stress/atom} keyword in LAMMPS does not represent the true heat current 
for systems with many-body potentials \cite{Surblys}.   Figure \ref{Figure2} shows, for the same simulation trajectory, the  time-dependent GK thermal conductivities for the $4\times 4\times4$ supercell 
 calculated using the {\it stress/atom} keyword alone for calculation of the heat current (gray curves) in comparison to the same quantities computed using the unfiltered hybrid current (blue). Both sets of curves correspond to unfiltered currents. It is obvious that the results for the two approaches are, indeed, not identical. In particular, the long-time value of $\kappa^{yy}(t)$ for the {\it stress/atom} keyword is
 about $10 \%$ lower than when the hybrid current is used, and  the long-time value of the off-diagonal component $\kappa^{xz}(t)$ is positive for the {\it stress/atom} keyword
 but negative when the hybrid current is used.
 Thus, the two heat-current definitions give noticeably different thermal conductivities for $\beta$-HMX 
  and, since the {\it stress/atom} keyword gives incorrect per-atom stress for systems 
  with many-body potentials, the hybrid heat current described in Sec. \ref{sd} should  be used to calculate thermal conductivity in such systems.
\subsection{Comparisons to other thermal conductivity results for HMX}
Hanson-Parr and Parr \cite{Parr} measured the thermal diffusivity and heat capacity of pressed-powder HMX for temperatures 293.15 K $\leq T \leq$ 433.15 K and obtained the thermal conductivity as a linear function of temperature by using that information in conjunction  with the sample densities. Their value of heat conductivity for  $T = 300$ K is 0.485 $\textrm{W\,m}^{-1}\,\textrm{K}^{-1}$. The values for pure (pressed) HMX taken from the Los Alamos Explosives Properties compendium \cite{Gibbs} are 0.502 $\textrm {W\,m}^{-1}\,\textrm{K}^{-1}$ at $T = 298.15$ K and 0.406 $\textrm {W\,m}^{-1}\,\textrm{K}^{-1}$ at 433.15 K. More recently, Dong et al. \cite{Dong} reported a value of 0.290 $\textrm {W\,m}^{-1}\,\textrm{K}^{-1}$ based on differential scanning calorimetry measurements. The thermal conductivity of HMX/Viton PBX formulations were measured as a function of HMX content \cite{Dobratz}. The results reveal increasing thermal conductivity with increasing HMX content; extrapolation to 100\% HMX yielded a thermal conductivity value of 0.519 $\textrm {W\,m}^{-1}\,\textrm{K}^{-1}$. In a recent study Lawless et al. \cite{Lawless} experimentally studied the room temperature thermal conductivity of the pressed-powder HMX at 90\% of the single crystal density and reported the value of 0.35 $\textrm {W\,m}^{-1}\,\textrm{K}^{-1}$. Our  values for $\bar\kappa({\infty})$  reported in Table \ref{t1} are higher than all experimental values. This can be provisionally attributed to fact that our results were obtained for a perfect single crystal whereas the experiments 
were performed on pressed powders. Thus, additional phonon scattering due to grain boundaries, dislocations, vacancies, and isotopic defects are present in experimental samples. All these additional scattering pathways decrease the thermal conductivity and can, in principle, be investigated using MD.

Numerical studies of thermal conductivities using the direct method \cite{Kroonblawd3,Chitsazi,Long,Perriot2} are usually performed for single crystals and reported for a given crystal direction specified by vector {\bf n}.
This vector  can be chosen to be of unit length, so that the scalar product  ${\bf n}\cdot{\bf n}=1$. Then, the scalar thermal conductivity along {\bf n}, $\kappa^{\bf n}$, can be obtained 
from the tensor $\kappa$ as 
\be
\kappa^{\bf n}=\bf n\cdot{\boldsymbol\kappa}\cdot \bf n. \label{direction}
\ee

Computational studies of the thermal conductivity of the $\beta$ and $\delta$ polymorphs of HMX were reported by 
Long et al. \cite{Long}, who performed MD simulations using the Smith-Bharadwaj \cite{Smith99} force field.  They applied a direct method in which a temperature gradient was imposed and the resulting heat current calculated. In the case of  $\beta$-HMX they used temperatures of 50 K and 380 K for the cold and hot regions of the crystal cell, respectively, to create the temperature gradient. This corresponded to the average temperature of 215 K. Unfortunately, Long et al. do not provide the system sizes for which their results were obtained, other than stating that they used less than 10000 atoms and that they found their results to be insensitive to the system size. For $\beta$-HMX ($P2_1/n$ space group setting), they reported values of 0.4718  $\textrm {W\,m}^{-1}\,\textrm{K}^{-1}$, 0.8008 $\textrm {W\,m}^{-1}\,\textrm{K}^{-1}$, and 0.6618 $\textrm {W\,m}^{-1}\,\textrm{K}^{-1}$ for conduction along the {\bf a}, {\bf b}, and {\bf c} crystal directions, respectively. The arithmetic average value, 0.6448  $\textrm {W\,m}^{-1}\,\textrm{K}^{-1}$, was taken to correspond to polycrystalline aggregates. This is an approximation as the rotational average of a symmetric second-rank tensor is given by one third of the trace of the tensor. Recall that whereas $\boldsymbol\kappa$ is a Cartesian tensor, $\beta$-HMX is monoclinic. In particular, lattice vectors {\bf{a}} and {\bf{c}} are non-orthogonal, such that, in general, 
$(\kappa^{\bf{a}}+\kappa^{\bf{b}}+\kappa^{\bf{c}})/3\neq(\kappa^{xx}+\kappa^{yy}+\kappa^{zz})/3$. However, given the small magnitude of 
$\kappa^{xz}$, the correction is relatively small.

The results of Long et al. \cite{Long} cannot be compared to ours directly  because our results were obtained for a different average temperature (300 K vs. 215 K) and because we observe a substantial size dependence of the thermal conductivity values. Generally, however, the thermal conductivity decreases with increasing temperature, and the  Long et al. result for conduction along the {\bf a} direction is lower than ours ($\kappa^{xx}$)  for all system sizes, including the extrapolation to infinite size, for both the  GK and HM 
analyses. Their results for conduction along the {\bf b} direction correspond to our $\kappa^{yy}$ and are approximately 8\% and 5\%
higher than our results for the infinite crystal obtained with the GK and HM approaches, respectively. The thermal conductivity along the  {\bf c} crystal direction for our conductivity tensor can be obtained by applying Eq. (\ref{direction}). For the infinite crystal the results are $0.61\pm0.05$ $\textrm {W\,m}^{-1}\,\textrm{K}^{-1}$ for GK and $0.66\pm0.07$ $\textrm {W\,m}^{-1}\,\textrm{K}^{-1}$ for HM. These numbers are very close to the corresponding $\kappa^{zz}$ values in Table \ref{t1}, which is not surprising in view of the very small 
$\kappa^{xz}$ values. Both of our values for the extrapolated thermal conductivity along the  {\bf c} crystal direction are slightly smaller than the one reported by Long et al. 

Chitsazi et al. \cite{Chitsazi} employed non-equilibrium MD to study the thermal conductivity of $\beta$-HMX at room temperature for three selected crystal directions. They employed a direct method in which a specified flux was imposed along the conduction direction, the resulting steady-state temperature gradient measured, and the thermal conductivity determined using Fourier’s law. They used the Smith-Bharadwaj \cite{Smith99,Bedrov3,Kroonblawd1} force field with the C-H  bonds constrained to their minimum bond-energy distance. The calculated thermal conductivity was rescaled to account for this constraint, using the approach described by Kroonblawd and Sewell \cite{KroonblawdSewell},  which was found to work well for the triclinic molecular
crystal 2,4,6-trinitrobenzene-1,3,5-triamine (TATB).
Chitsazi et al. considered thermal conduction along crystal directions normal to (011), (110), and (010) crystal planes. Three system sizes were considered  for each direction with the infinite size limits  calculated using Matthiessen’s rule. Two slightly different numerical protocols, that they called different flux (DF) and single flux (SF), were
applied leading to two sets of thermal conductivity values for the three directions studied. 
 Their infinite-size-limit thermal conductivity values
are listed in Table \ref{t2}, where we 
 compare $\boldsymbol\kappa({\infty})$ tensors to the results of the present study by applying Eq. (\ref{direction}) to our $\boldsymbol\kappa({\infty})$ GK and HM tensors for the crystal directions considered by Chitsazi et al.
%The results are shown in Table \ref{t2}.
%to obtain for the direction normal to (011): 0.65 $\textrm {W\,m}^{-1}\,\textrm{K}^{-1}$ (GK) and 0.69 $\textrm {W\,m}^{-1}\,\textrm{K}^{-1}$ (HM),
%for the direction normal to (110): 0.79 $\textrm {W\,m}^{-1}\,\textrm{K}^{-1}$ (GK) and 0.83 $\textrm {W\,m}^{-1}\,\textrm{K}^{-1}$ (HM), and for the direction normal to (010), which is the direction along the {\bf b} crystal axis:
%0.74 $\textrm {W\,m}^{-1}\,\textrm{K}^{-1}$ (GK) and 0.76 $\textrm {W\,m}^{-1}\,\textrm{K}^{-1}$ (HM). 
\begin{table*}
\caption{Thermal conductivities of  $\beta$-HMX at (300 K, 1 atm). Units are $\textrm {W\,m}^{-1}\,\textrm{K}^{-1}$  for three crystal directions obtained in this work compared to the results of Chitsazi et al. \cite{Chitsazi}.}
\centering
\begin{tabular}{ccccccc}
%\toprule
	&{\textrm{Normal to (011)}}	&{\textrm{Normal to (110)}} &{\textrm{Normal to (010)}} \\
\hline
\textrm{This work}		&$0.65\pm0.04$ (GK)      &$0.79\pm0.05$ (GK)   &$0.74\pm0.06$ (GK)   \\
      &$0.69\pm0.05$ (HM)  &$0.83\pm0.07$ (HM)& $0.76\pm0.08$ (HM) \\
      \hline
\textrm{Chitsazi et al.}		& $0.553\pm0.050$ (DF) &$ 0.701\pm0.090$ (DF)  & $0.632\pm0.070$ (DF) \\
		&$0.557$ (SF)  &$ 0.671\pm0.020 $ (SF) & $0.665\pm0.020$ (SF)\\
\hline
\end{tabular}
\label{t2} 
\end{table*}
Our values are approximately 20\% to 25\% higher than the ones reported by Chitsazi et al. We do not have a definite explanation for this discrepancy.
It is possible that the difference stems from the fact that the direct method, which involves a temperature gradient along the conduction direction, gives effective conductivity 
of the thermally inhomogeneous system, which can be different from the conductivity of the same system at homogeneous thermal equilibrium. It is also possible that freezing C-H bonds in Ref. \cite{Chitsazi} has a stronger effect  on the thermal conductivity for $\beta$-HMX than TATB, that is not
fully accounted for by simple rescaling as was 
done by Chitsazi et al. based on the success of the same approach  for TATB \cite{KroonblawdSewell}. In a topically adjacent study, Algaer and M\"{u}ller-Plathe \cite{Algaer} used velocity-reversal non-equilibrium MD to study thermal conductivity anisotropy in the syndiotactic $\delta$-polymorph of crystalline polystyrene. The work included a sub-study of sensitivity of predicted conductivities to six different bond-constraint combinations that ranged from fully flexible chains to globally constrained C-H bonds to various constraint patterns among the phenyl group C-C and/or C-H bonds to a pattern in which fully bond-constrained styryl moieties were separated by flexible C-C linker bonds. They reported a weak trend toward decreased conductivity with increasing numbers of bond constraints in the dynamics, albeit in an “erratic” fashion that was not monotonic with the number of constraints and did not yield a simple interpretation of cause and effect, beyond the intuitive observation that constrained C-C bonds in the backbone led to the lowest conductivities among (most of) the constraint-pattern / conduction-direction cases examined.

Very recently, Perriot and Cawkwell \cite{Perriot2} reported a thorough study of the temperature- and pressure-dependent thermal conductivity tensor of 
$\beta$-HMX, for 200 K 
$\leq T \leq 500$ K and $0\,\,\,{\textrm{GPa}} \leq P \leq 5$ GPa. (We draw the reader’s attention to Fig. 7 of their paper, which effectively summarizes the published experimental and theoretical thermal conductivity data for HMX  for $P = 0$ GPa.) The predictions are based on a direct, non-equilibrium velocity-exchange MD approach that was applied for all conduction directions required to determine the tensor at a given ($T$, $P$) state. Finite-size effects were accounted for using Mathiessen’s rule. Broadly speaking, the present values for the tensor elements are decidedly larger than those due to Cawkwell and Perriot at standard ambient conditions. Our result for $\bar\kappa({\infty})$  and that of Cawkwell and Perriot at (300 K, 0 GPa) differ by almost 40\%, with the Chitsazi et al. \cite{Chitsazi} result falling approximately midway between the others. This is concerning as, although the three sets of authors all used different approaches, all three studies are thought to have used very nearly the same force field, the main known distinctions being: (1) use of C-H bond constraints by Chitsazi et al.; (2) different N-O and, in the present study, C-H bond force constants by Sewell and co-workers vs. Perriot and Cawkwell; and (3) incorporation of the steep, very-short-range non-bonded interatomic potential in the present study. Perriot and Cawkwell discuss potential implications of the first two points of distinction in some detail. Regarding the third, given the very short distance interval for which the steep repulsive core is practically non-zero, we are confident that it has little if any effect on the present predictions. The explanation of why ostensibly equivalent approaches, simulated using such similar force fields, yields such different results for the thermal conductivity is an outstanding question that deserves further attention.

%\begin{table}
%\caption{Thermal conductivity tensor components of  $\beta$-HMX for different sample sizes obtained from the HM approach}
%\centering
%\begin{tabular}{cccccc}
%\toprule
%\textrm{Cell size}	&$\kappa^{xx}$	&$\kappa^{yy} $&$\kappa^{zz}$&$\kappa^{xz}$ \\
%\hline
%$4\times4\times4$		& $0.58\pm0.03$ &$ 0.65\pm0.04 $ & $0.45\pm0.02$  & $-0.02\pm0.01$\\
%$5\times5\times5$		&$0.63\pm0.02$      &$0.69\pm0.03$  &$0.49\pm0.02$& $-0.01\pm0.01$  \\
%$6\times6\times6$		&$0.67\pm0.04$  &$0.69\pm0.03$ &$0.53\pm0.02$& $-0.02\pm0.01$ \\
%$8\times8\times8$	&$0.70\pm0.02$  &$0.70\pm0.03$& $0.52\pm0.02$& $-0.02\pm0.01$\\
%$\infty$	&$0.87\pm0.09$  &$0.76\pm0.08$& $0.65\pm0.07$ & $-0.01\pm0.01$ \\
%\hline
%\end{tabular}
%\label{t2}
%\end{table}

%\begin{table}
%\caption{Thermal conductivity tensor components of  $\beta$-HMX for different sample sizes obtained from the GK approach using double-exponential fitting.}
%\centering
%%% \tablesize{} %% You can specify the fontsize here, e.g., \tablesize{\footnotesize}. If commented out \small will be used.
%\begin{tabular}{cccccc}
%\toprule
%\textrm{Cell size}	&$\kappa^{xx}$	&$\kappa^{yy} $&$\kappa^{zz}$ \\
%
%$4\times4\times4$		& $0.58$ &$ 0.65 $ & $0.45$  \\
%$5\times5\times5$		&$0.63$      &$0.69$  &$0.49$  \\
%$6\times6\times6$		&$0.67$  &$0.69$ &$0.53$ \\
%$8\times8\times8$	&$0.70$  &$0.70$& $0.52$\\
%\textrm{Infinite size}	&$0$  &$0$& $0$\\
%\end{tabular}
%\label{300}
%\end{table}

 \section{Conclusions}
  Heat-current correlation functions required for equilibrium MD-based Green-Kubo calculations of thermal conductivity in crystals exhibit large-amplitude oscillations, even for time-converged simulations and especially for polyatomic crystals. In this work, we showed that some contributions to the heat current—namely those which correspond to the time derivative of a bounded linear function of atomic positions and therefore do not contribute to the thermal conductivity—can be filtered from the raw heat-current data in multiple physically justified ways. The HCCFs obtained from the filtered current are far less oscillatory than for the unfiltered current, and the resulting GK time-dependent thermal conductivities are far smoother numerical functions than when the unfiltered current is used. This latter fact motivates an approach for estimating the time-asymptotic conductivity, wherein a physically motivated bi-exponential function is fitted to the time-dependent conductivities and extrapolated to infinite time. The combination of filtered heat current with fitting the resulting time-dependent conductivity directly (as opposed to fitting the HCCF to a much more complicated function that is subsequently integrated over time) appears to provide a simple, relatively well-posed approach for extracting crystal thermal conductivity from Green-Kubo simulations. Although we do not develop it here, the same basic filtering concepts can, in principle, be extended to filter quadratic and higher-order terms in atomic displacements and velocities that also correspond to time derivatives of bounded functions and therefore do not contribute the thermal conductivity. Doing so would further reduce heat-current oscillations that are not relevant for obtaining the thermal conductivity. 

The preceding concepts were applied to predictions of the thermal conductivity of the monoclinic polyatomic molecular crystal $\beta$-HMX at standard ambient conditions. The results were obtained based on 120 ns of accumulated MD simulation time for each of four 3D-periodic crystal supercell sizes ranging between 3584 and 28672 atoms. Extrapolations to infinite crystal size were performed using Matthiesen’s rule. The GK and HM predictions for the thermal conductivity tensor obtained using the filtered current and (for GK) the bi-exponential fitting approach were found to be in good agreement. The predicted $\beta$-HMX conductivity values are all similar to, but larger than, most of the comparable values in the literature, including MD predictions based on very nearly the same force field as employed here but obtained using different theoretical approaches. The source of this discrepancy, which has been observed for GK versus direct approaches, and indeed among different direct approaches, is an area of ongoing investigation.

 \section*{Supplementary material} \label{sm}
 The supplementary material contains results for the $xy$, $yx$, $yz$, and $zy$ components of the HCCFs and thermal conductivity tensors, a description of the error analysis, and sample LAMMPS input decks.

\section*{Acknowledgments}
This research was funded by Air Force Office of Scientific Research, grant number FA9550-19-1-0318. The authors are 
grateful for an AFOSR-DURIP equipment grant for computing resources, grant number FA9550-20-1-0205.
%\end{acknowledgments}
\section*{Declaration of Competing Interest}
The authors have no conflicts to disclose.
%\section*{Data Availability}
%The data that support the findings of this study are available from the authors upon %reasonable request.

%merlin.mbs apsrev4-1.bst 2010-07-25 4.21a (PWD, AO, DPC) hacked
%Control: key (0)
%Control: author (8) initials jnrlst
%Control: editor formatted (1) identically to author
%Control: production of article title (-1) disabled
%Control: page (0) single
%Control: year (1) truncated
%Control: production of eprint (0) enabled

%%

 \bibliographystyle{elsarticle-num} 
 %\bibliography{HMXHEAT}

%% else use the following coding to input the bibitems directly in the
%% TeX file.

\end{document}